\documentclass[12pt,a4paper]{article}

\newcommand{\rdot}{{\dot{\rho}}}
\newcommand{\sdot}{{\dot{\sigma}}}

\begin{document}

\title{Quantum gravity: the Inverse problem}

\author{Ll. Bel\thanks{e-mail:  wtpbedil@lg.ehu.es}}
\maketitle

\date{}

\begin{abstract}
Quantizing the gravitational field described by General relativity being a notorious difficult, unsolved and maybe meaningless problem I use in this essay a different strategy: I consider a linear theory in the framework of Special relativity where the potentials are the components of four linear forms. General relativity and other similar covariant non linear theories can be formulated in this way. The theory that I propose is Lorentz invariant, linear,  simple, and can be quantized following similar steps to those that led to quantum electrodynamics.
\end{abstract}

\section{Gravity theory}

{\it Field potentials}
\vspace{1cm}

I use Greek indices and doted Greek indices. Both can be raised or lowered with the Minkowski metric $\eta_{\alpha\beta}$; and indices $\rho$ and $\rdot$ can be contracted as usual:

\begin{equation}
\label{indices}
\alpha,\beta\cdots\mu=0,1,2,3 \quad \rdot,\sdot\cdots=0,1,2,3
\end{equation}

Let us consider four linear forms $\theta^\rdot_\alpha$, functions of $x^\alpha$, named the potentials, in the framework of Special relativity. The field components are:

\vspace{1cm}
{\it Field components}
\vspace{1cm}

\begin{equation}
\label{Field components}
F^\rdot_{\alpha\beta}=\partial_\alpha \theta^\rdot_{\beta} -\partial_\beta \theta^\rdot_{\alpha}, \quad F_\alpha=F^\rdot_{\alpha\rho}
\end{equation}
They are invariant under the local gage transformations:

\vspace{1cm}

\begin{equation}
\label{gage}
\theta^\rdot_{\alpha} \mapsto \theta^\rdot_{\alpha}+\partial_\alpha \zeta^\rdot
\end{equation}
From its definition it follows that:

\begin{equation}
\label{structure eqs}
\partial_{\alpha}F^\rdot_{\beta\gamma}+\partial_{\beta}F^\rdot_{\gamma\alpha}+\partial_{\gamma}F^\rdot_{\alpha\beta}=0
\end{equation}

{\it Lagrangian}
\vspace{1cm}

Following the steps of Maxwell theory the Lagrangian I am interested in is:

\begin{equation}
\label{Lagrangian}
{\cal L}=-\frac14 F^\rdot_{\alpha\beta} F^\sdot_{\lambda\mu}\eta^{\alpha\lambda}\eta^{\beta\mu}\eta_{\rho\sigma}
+\frac12 F^\rdot_{\alpha\rdot}F^\sdot_{\lambda\sdot}\eta^{\alpha\lambda}
\end{equation}
this particular choice being justified later.

This Lagrangian is globally Lorentz invariant, and locally gage invariant. The field equations derived from it are:

\begin{equation}
\label{Field equations}
G^\beta_\rdot=j^\beta_\rdot
\end{equation}
$j^\beta_\rdot$ being the four conserved currents one wishes to consider as sources, and

\begin{equation}
\label{Gtensor}
G^\beta_\rdot= \partial_\alpha F^{\alpha\beta}_\rdot-\partial_\alpha F^\alpha \delta^\beta_\rho+\partial_\rho F^\beta,
 \quad F_\alpha=F^\rdot_{\alpha\rho}
\end{equation}
is a conserved tensor:

\begin{equation}
\label{conservation}
\partial_\beta G^\beta_\rdot\equiv 0
\end{equation}

{\it Metric of space-temps}
\vspace{1cm}

The tetrad formalism of General relativity starts also with four linear forms $\theta^\rdot_\alpha$  but instead of considering the field variables (\ref{Field components}) it uses them to introduce the  4-dimensional Riemannian hyperbolic metric:

\begin{equation}
\label{metric}
g_{\alpha\beta}=\eta_{\rdot\sdot}\theta^\rdot_\alpha \theta^\sdot_\beta
\end{equation}
and uses as field equations the highly non linear Einstein's equations, that beautiful and successful as they are  at the macroscopic level, they are stubbornly resisting quantization.

This metric is locally Lorentz covariant:
\begin{equation}
\label{Lorentz}
\theta^{\rdot\prime}_\alpha=L^{\rdot^\prime}_\sdot \theta^b_\alpha\Rightarrow
g^\prime_{\alpha\beta}=g_{\alpha\beta}
\end{equation}
locally meaning  that the matrix elements $L^{\rdot^\prime}_\sdot$ could be functions of $x^\alpha$. But it is not gage invariant. So that to each solution of the field equations (\ref{Field equations}) will correspond a  functional family of metrics $g_{\alpha\beta}(x^\alpha,S^\rdot)$

On the other hand the Weitzenb{\"o}ck formalism starts with a Riemannian hyperbolic metric but defines the potentials by a diagonal decomposition (\ref{metric}) and uses as main concept that of torsion instead of that of curvature, \cite{Weitzenbock}-\cite{Maluf}.

\vspace{1cm}

Let us assume that:

\begin{equation}
\label{thetas}
\theta^\rdot_\alpha = \delta^\rdot_\alpha+\frac12 {\hat f}^\rdot_{\alpha}, \quad {\hat f}^\rdot_{\alpha}=O(1)
\end{equation}
and:

\begin{equation}
\label{symmetry}
\eta_{\alpha\rho}{\hat f}^\rdot_\beta-\eta_{\beta\rho}{\hat f}^\rdot_\alpha=0
\end{equation}
where ${\hat f}^\rdot_{\alpha}$ are small quantities so that its powers can be neglected. The corresponding metric will be:

\begin{equation}
\label{approximate metric}
g_{\alpha\beta}=\eta_{\alpha\beta}+h_{\alpha\beta},
\end{equation}
where:

\begin{equation}
\label{h's}
h_{\alpha\beta}=\frac12(\eta_{\alpha\rho}{\hat f}^\rdot_\beta+\eta_{\beta\rho}{\hat f}^\rdot_\alpha)
\end{equation}
that with a gage transformation (\ref{gage}) will become:

\begin{equation}
\label{approximate gage}
g_{\alpha\beta}=\eta_{\alpha\beta}+h_{\alpha\beta}+\frac12(\partial_\alpha \zeta_\beta+\partial_\beta \zeta_\alpha),
\quad \zeta_\alpha=\eta_{\alpha\sdot}\zeta^\sdot_\alpha
\end{equation}

A straightforward calculation shows that:

\begin{equation}
\label{S=G}
S_{\alpha\beta}=-\frac12 G_{\alpha\beta}
\end{equation}
where $G_{\alpha\beta}$ is the tensor defined in (\ref{Gtensor}) and $S_{\alpha\beta}$ is the Einstein  tensor of the linear metric (\ref{approximate metric}):

\begin{equation}
\label{Linear Einstein}
S_{\alpha\beta}=R_{\alpha\beta}-\frac12 R\eta_{\alpha\beta}
\end{equation}
where:

\begin{equation}
R_{\alpha\lambda}=-\frac12\eta^{\beta\mu}(\partial_{\alpha\lambda}h_{\beta\mu}+\partial_{\beta\mu}h_{\alpha\lambda}
-\partial_{\alpha\mu}h_{\beta\lambda}-\partial_{\beta\lambda}h_{\alpha\mu})
\end{equation}

Therefore a linear new theory based on the Lagrangian (\ref{Lagrangian}) is equivalent to Einstein's theory when both are considered at the linear approximation.

\section{Free Graviton waves}

A graviton wave is by definition a solution of the vacuum field equations:

\begin{equation}
\label{vacuum}
G_{\alpha\beta}=0
\end{equation}
with:

\begin{equation}
\label{waves}
\theta_\alpha^\rdot=f_\alpha^\rdot\exp(i\, l_{\sigma}x^\sigma), \quad l_\sigma l^\sigma=0
\end{equation}
the propagation vector $l_\sigma$ being a null vector and the polarization tensor $f_\alpha^\rdot$ a constant tensor that by a gage transformation
(\ref{gage}) where:

\begin{equation}\label{zeta}
\zeta^\rdot=\lambda^\rdot \exp(i\,l_\sigma x^\sigma)
\end{equation}
becomes:

\begin{equation}\label{Gage fixing}
f^\rdot_\alpha \rightarrow f^\rdot_\alpha + l_\alpha \lambda^\rdot
\end{equation}

The gage invariant field components are thus:

\begin{equation}
\label{wave fields}
F_{\alpha\beta}^\rdot=i(f_\beta^\rdot l_\alpha - f_\alpha^\rdot l_\beta)\exp(i\, l_{\sigma}x^\sigma),
\end{equation}
and:

\begin{equation}
\label{wave contraction}
F_\alpha=i(f l_\alpha - f_\alpha^\rdot l_\rho)\exp(i l_{\sigma}x^\sigma), \quad f=f_\rho^\rdot
\end{equation}

The vacuum field equations are thus:

\begin{equation}
\label{polarization 1}
-l^\alpha f_\alpha^\rdot l_\beta+(l^\alpha f_\alpha^\sdot l_\sigma)\delta_\beta^\rho+l^\rho(f l_\beta-f_\beta^\sdot l_\sigma)=0
\end{equation}
The contraction of the two indices $\beta$ and $\rdot$ yields:

\begin{equation}
\label{scalar equation}
l^\alpha f_\alpha^\rdot l_\rho=0
\end{equation}
and therefore (\ref{polarization 1}) becomes:

\begin{equation}
\label{polarization 2}
-l^\alpha f_\alpha^\rdot l_\beta+f l_\beta l^\rho-l^\rho f_\beta^\sdot l_\sigma=0
\end{equation}
Introducing the symmetric and antisymmetric parts of  $f_{\alpha\beta}=\eta_{\rho\beta} f^\rdot_\alpha$:

\begin{equation}
\label{sym-ant def}
f^{-}_{\alpha\beta}=\frac12(f_{\alpha\beta}-f_{\beta\alpha}), \quad
f^{+}_{\alpha\beta}=\frac12(f_{\alpha\beta}+f_{\beta\alpha})
\end{equation}
so that:
\begin{equation}\label{fab}
 f_{\alpha\beta}= f^{+}_{\alpha\beta}+ f^{-}_{\alpha\beta}
\end{equation}
and defining:

\begin{equation}
\label{vector v}
v^{-}_\rho\equiv l^\alpha f^{-}_{\alpha\rho}, \quad v^{+}_\rho\equiv l^\alpha f^{+}_{\alpha\rho}
\end{equation}
the above equation (\ref{polarization 2}) becomes:

\begin{eqnarray}
\label{sym-ant eq1}
 -(v^{+}_\beta l_\rho+v^{+}_\rho l_\beta)+ f l_\beta l_\rho &+& \\
 -(v^{-}_\beta l_\rho-v^{-}_\rho l_\beta) &=&0
\end{eqnarray}
The first row is symmetric in $\beta$ and $\rho$ and the second row is antisymmetric and therefore the equation is equivalent to the two equations system:

\begin{eqnarray}
\label{sym-ant eq2}
 -(v^{+}_\beta l_\rho+v^{+}_\rho l_\beta)+ f\, l_\beta l_\rho &=&0 \\
 -(v_\beta l_\rho-v_\rho l_\beta)  &=&0
\end{eqnarray}
from where it follows that
\begin{equation}
\label{l+,l-}
l^\alpha f^{-}_{\alpha\rho}=a^{-} l_\rho,  \quad l^\alpha f^{+}_{\alpha\rho}=\frac12 f l_\rho, \quad f=f_\rho^\rdot
\end{equation}
Using (\ref{fab}) leads the two conditions above to the more convenient form:

\begin{equation}
\label{b1,b2}
l^\alpha f_{\alpha\rho}=b_1 l_\rho,  \quad l^\alpha f_{\rho\alpha}= b_2 l_\rho,
\end{equation}
with:

\begin{equation}\label{f}
 b_1+b_2=f \ ;
\end{equation}
\vspace{1cm}
$b_1$ is gage invariant, but $b_2$ is not.

Let $u^\alpha$ be a time-like unit vector, to be named a gage-fixing vector. With a gage transformation (\ref{Gage fixing}) we get:

\begin{equation}\label{e0}
u^\alpha f_{\alpha\beta}\rightarrow u^\alpha f_{\alpha\beta}+(u^\alpha l_\alpha) \zeta_\beta
\end{equation}
and therefore, since the coefficient of $\zeta_\beta$ is different of zero,it is always possible to implement the condition:

\begin{equation}\label{g-codition}
u^\alpha f_{\alpha\beta}=0
\end{equation}
Doing that implies that a gage invariant translation of (\ref{b1,b2}) is:

\begin{equation}\label{GageInv}
 f^\rdot_{\alpha\beta}l^\alpha=b_1 \, l^\rdot l_\beta,\quad f^\rdot_{\alpha\beta}l_\rho=0
\end{equation}

\section{Helicities}

Let $\vec{e}_0$ be any time-like, gage-fixing unit vector; $\vec{e}_1$ a unit-vector on the 2-plane $\Pi$ defined by $\vec{e}_0$ and $\vec l$, and complete an orthogonal frame with two unit orthogonal vectors $\vec{e}_a$ (a=2,3) on the 2-plane orthogonal to $\Pi$.

Using the  conditions (\ref{GageInv}), and the above defined reference frame, the strict matrix components of $f_{\alpha\beta}$ are:

\vspace{0.5cm}

\hspace{3cm}\begin{tabular}{cccc}
$f_{00}$ & $-f_{00}$ & $f_{02}$ & $f_{03}$\\
$-f_{00}-b_1$ & $f_{00}+b_1$ & $-f_{02}$ & $-f_{03}$\\
$f_{20}$ & $f_{20}$ & $f_{22}$ & $f_{23}$\\
$f_{30}$ & $f_{30}$ & $f_{32}$ & $f_{33}$\\
\end{tabular}

\vspace{0.5cm}
\hspace{-0.5cm} With this simplification the preceding matrix becomes:

\vspace{0.5cm}

\hspace{3cm}\begin{tabular}{cccc}
$0$ & $0$ & $0$ & $0$\\
$-b_1$ & $b_1$ & $0$& $0$\\
$f_{20}$ & $f_{20}$ & $f_{22}$ & $f_{23}$\\
$f_{30}$ & $f_{30}$ & $f_{32}$ & $f_{33}$\\
\end{tabular}

\vspace{0.5cm}
\hspace{-0.5cm} Now two cases might be considered. If we assume that the matrix $f_{\alpha\beta}$ is symmetric then the matrix becomes:

\vspace{0.5cm}

\hspace{3cm}\begin{tabular}{cccc}
$0$ & $0$ & $0$ & $0$\\
$0$ & $0$ & $0$& $0$\\
$0$ & $0$ & $f_+$ & $f_\times$\\
$0$ & $0$ & $f_\times$ & $-f_+$\\
\end{tabular}

\vspace{0.5cm}
\hspace{-0.5cm} This corresponds to an helicity 2 of the graviton.

If we assume that the matrix $f_{\alpha\beta}$ is antisymmetric then the matrix becomes:

\vspace{0.5cm}

\hspace{3cm}\begin{tabular}{cccc}
$0$ & $0$ & $0$ & $0$\\
$0$ & $0$ & $0$& $0$\\
$0$ & $0$ & $0$ & $\bar f$\\
$0$ & $0$ & $-\bar f$ & $0$\\
\end{tabular}

\vspace{0.5cm}
\hspace{-0.5cm} that corresponds to an helicity 0.

I recover thus the algebraic structure of the graviton concept we are familiar with from the framework of General relativity at the linear approximation, (\cite{Weinberg}), (\cite{Kiefer})

\section{Canonical energy-momentum tensor}

The canonical energy-momentum tensor:

\begin{equation}\label{Canonical}
t^\alpha_\gamma=\partial_\gamma \theta^\rdot_\delta\frac{\partial \cal L}{\partial(\partial_\alpha \theta^\rdot_\delta)}-{\cal L}\delta^\alpha_\gamma
\end{equation}
corresponding to the Lagrangian (\ref{Lagrangian}) can be easily calculated using the following convenient form of the derivatives of ${\cal L}$ :

\begin{eqnarray}\label{Der Lagrangian}
\frac{\partial {\cal L}}{\partial(\partial_\alpha\theta^\rdot_\beta)}=
-\frac12\partial_\lambda\theta^\sdot_\mu(&(&\eta^{\alpha\lambda}\eta^{\beta\mu}-\eta^{\beta\lambda}\eta^{\alpha\mu})\eta_{\rho\sigma}\\
-&(&\eta^{\alpha\lambda}\delta^\beta_\rho-\eta^{\beta\lambda}\delta^\alpha_\rho)\delta^\mu_\sigma
+(\eta^{\alpha\mu}\delta^\beta_\rho-\eta^{\beta\mu}\delta^\alpha_\rho)\delta^\lambda_\sigma))
\end{eqnarray}

The preceding result, with the corresponding expression of the Lagrangian (\ref{Lagrangian}) and the use of (\ref{GageInv}) proves, after a simple calculation, that for a graviton field the Canonical energy-momentum tensor is:

\begin{equation}\label{e-m}
t^\alpha_\gamma=-\frac12(f^\rdot_\beta f^\sdot_\mu \eta^{\beta\mu} \eta_{\rho\sigma}+2b_1 b_2)\exp(2 \,i\, l_\rho x^\rho)l^\alpha l_\gamma
\end{equation}
that can be simplified with a gage transformation that makes $b_2=0$.

\section{Quantum gravity Lagrangian}

This essay is a proposal to reverse the problem of Quantum gravity: instead of accepting that General relativity is a theory that applies both at the macroscopic level and the microscopic one, I suggest that we should keep General relativity to deal with macroscopic problems only and try to find a new theory to deal with microscopic ones, with the condition that at the linear approximation both theories coincide.

Given four linear forms $\theta^\rdot_\alpha$ there are two equivalent descriptions of the geometrical frame of General relativity. The best known and almost universally used is based on the concept of Riemann curvature, while a second one, little used, is based on the concept of Weitzenb\"{o}ck torsion \cite{Schucking}-\cite{Bel}. The formula (\ref{S=G}) above proves this equivalence at the first approximation. The unrestricted equivalence is discussed in \cite{Bel}.

To be more specific: I propose to discuss the quantum interaction of a spin-1/2 wave function $\psi$  with a Poincar\'{e} invariant theory of the gravitational field $F^\rdot_{\alpha\beta}$ based on the Lagrangian:

\begin{equation}
\label{QLagrangian}
{\cal L}=-\frac14 F^\rdot_{\alpha\beta} F^\sdot_{\lambda\mu}\eta^{\alpha\lambda}\eta^{\beta\mu}\eta_{\rdot\sdot}
+\frac12 F^\rdot_{\alpha\rdot}F^\sdot_{\lambda\sdot}\eta^{\alpha\lambda}
+i{\bar\Psi}\gamma^\mu\partial_\mu\Psi-m{\bar\Psi}\Psi-p_\rho\bar{\Psi}\gamma^\mu\Psi \theta^\rdot_{\mu}
\end{equation}
where $p_\rho$ are the components of the 4-momentum of the interacting fermion with mass $m$ so that:

\begin{equation}
\label{mass}
\eta_{\rho\sigma}q^\rho q^\sigma=-m^2
\end{equation}

The equations of motion that follow are:

\begin{eqnarray}
\label{EqF}
\partial_\alpha F^{\alpha\beta}_\rdot-\partial_\alpha F^\alpha \delta^\beta_\rho+\partial_\rho F^\beta&=&q_\rho{\bar\Psi}\gamma^\beta\Psi, \nonumber \\
i\gamma^\mu\partial_\mu\Psi-m\Psi~&=&q_\rho\gamma^\mu\Psi \theta^\rdot_{\mu}
\end{eqnarray}

They are invariant under local gage transformations:

\begin{equation}
\label{quantum gage}
\theta^\rdot_\alpha\rightarrow \theta^\rdot_\alpha+\partial_\alpha \zeta^\rdot, \quad \psi\rightarrow \exp(-i q_\sigma \zeta^\sdot)\psi
\end{equation}
and global Lorentz transformations.

Quantization can now proceed as usual introducing the operators:

\begin{eqnarray}\label{operator f(+2)}
 \hat f^+_{\alpha\rho}(x^\beta,p_\mu)=\sum_{\sigma=\pm 2}\int \frac{d^3 l}{\sqrt{2|\vec l|}}
(&a(\vec l,p_\mu,\sigma)e_{\alpha\rho}(\vec l,p_\mu,\sigma)\exp(i\,l_\beta x^\beta) \nonumber \\
 + &a^\dagger(\vec l,p_\mu,\sigma)e^*_{\alpha\rho}(\vec l,p_\mu,\sigma)\exp(-i\,l_\beta x^\beta))
\end{eqnarray}

where to describe the interaction of a fermion with a graviton with helicity 2:

\begin{eqnarray}\label{e's}
e_{\alpha\rho,+2}(\vec l,p_\mu,+2)=e_{2\alpha}e_{2\rho}-e_{3\alpha}e_{3\rho} \\
e_{\alpha\rho}(\vec l,p_\mu,-2)=e_{2\alpha}e_{3\rho}+e_{3\alpha}e_{2\rho}
\end{eqnarray}
where $e_{2\alpha}$ and $e_{3\alpha}$ are any two unit complex vectors  orthogonal to $l_\alpha$ and $p_\alpha$.

The formulas corresponding to helicity 0 are:

\begin{eqnarray}\label{operator f(0)}
 \hat f^-_{\alpha\rho}(x^\beta,p_\mu)=\int \frac{d^3 l}{\sqrt{2|\vec l|}}
(&a(\vec l,p_\mu)e_{\alpha\rho}(\vec l,p_\mu)\exp(i\,l_\beta x^\beta) \nonumber \\
 + &a^\dagger(\vec l,p_\mu)e^*_{\alpha\rho}(\vec l,p_\mu)\exp(-i\,l_\beta x^\beta))
\end{eqnarray}

where:

\begin{equation}\label{e's}
e_{\alpha\rho}(\vec l,p_\mu)=e_{2\alpha}e_{3\rho}-e_{3\alpha}e_{2\rho}
\end{equation}

Notice that considering  the creation, $a$, and annihilation ,$a\dagger$, amplitudes of a graviton as explicit functions of the 4-momentum $p_\mu$ of the Dirac particle that is the real vector "charge" of the gravitational field makes of (\ref{operator f+}) and (\ref{operator f-}) truly tensor definitions.
(compare with \cite{Weinberg} and Kiefer)

This dependence should be reminded also when writing the corresponding non zero commutators:

\begin{equation}\label{commutators+}
 [a(\vec l, p_\mu, \sigma),a^\dagger(\vec l^\prime, p_\mu, \sigma^\prime)]=\delta_{\sigma\sigma^\prime}\delta(\vec l-\vec l^\prime),
\end{equation}
or:

\begin{equation}\label{commutators+}
 [a(\vec l, p_\mu),a^\dagger(\vec l^\prime, p_\mu)]=\delta(\vec l-\vec l^\prime),
\end{equation}
depending on the case.

General relativity is formulated in terms of a curved 4-dimensional geometry, using the concepts of curvature or torsion depending on taste \cite{Weitzenbock}-\cite{Maluf}. But another conceptual ingredient in the theory is that schematized observers are part of the theory and this make sense only at the macroscopic level.

In my opinion Quantum gravity could be based on Eqs. (\ref{EqF}) that does not depend on Riemann curvature, or Weitzenb\"{o}ck torsion of space-time.
This can be accepted  at the microscopic level, where the main concepts are those of gravitons mediating the gravitational interactions of Dirac particles, and observers are no more part of the theory. Observers are only preparing experiments and observing results as they always have done when dealing with photons and Dirac particles.


\end{document}